\documentclass[pdflatex,sn-aps]{sn-jnl}

\usepackage{graphicx}%
\usepackage{multirow}%
\usepackage{amsmath,amssymb,amsfonts}%
\usepackage{amsthm}%
\usepackage{mathrsfs}%
\usepackage[title]{appendix}%
\usepackage{xcolor}%
\usepackage{textcomp}%
\usepackage{booktabs}%
\usepackage{algorithm}%
\usepackage{algorithmicx}%
\usepackage{algpseudocode}%
\usepackage{listings}%
\usepackage[normalem]{ulem}%
\usepackage{booktabs}%
\usepackage{longtable}
\usepackage{subcaption}
\usepackage{adjustbox}
\usepackage{array}
\usepackage{url} 
\usepackage{float}
\usepackage[section]{placeins}
\usepackage{colortbl}
\usepackage[normalem]{ulem}

\begin{document}
\title[Article Title]{Causal Modelling of Cryptocurrency Price Movements Using Discretisation-Aware Bayesian Networks}

\author[1]{Rasoul Amirzadeh\thanks{Corresponding author: rasoul.amirzadeeh@gmail.com}}
\author[1]{Dhananjay Thiruvady}
\author[2]{Asef Nazari}
\author[1]{Mong Shan Ee}

\affil[1]{School of Information Technology, Deakin University,   Australia}
\affil[2]{Deakin Business School, Deakin University,  Australia}

\abstract{
This study identifies the key factors influencing the price movements of major cryptocurrencies, Bitcoin, Binance Coin, Ethereum, Litecoin, Ripple, and Tether, using Bayesian networks~(BNs). This study addresses two key challenges: modelling price movements in highly volatile cryptocurrency markets and enhancing predictive performance through discretisation-aware Bayesian Networks. It analyses both macro-financial indicators (gold, oil, MSCI, S\&P 500, USDX) and social media signals (tweet volume) as potential price drivers. Moreover, since discretisation is a critical step in the effectiveness of~BNs, we implement a structured procedure to build 54~BNs models by combining three discretisation methods (equal interval, equal quantile, and k-means) with several bin counts. These models are evaluated using four metrics, including balanced accuracy, F1 score, area under the ROC curve and a composite score. Results show that equal interval with two bins consistently yields the best predictive performance. We also provide deeper insights into each network’s structure through inference, sensitivity, and influence strength analyses. These analyses reveal distinct price-driving patterns for each cryptocurrency, underscore the importance of coin-specific analysis, and demonstrate the value of BNs for interpretable causal modelling in volatile cryptocurrency markets.}


\maketitle

\section{Introduction}
\begin{figure*}[!htbp]
  \centering \includegraphics[width=\linewidth,height=0.3\textheight, keepaspectratio]{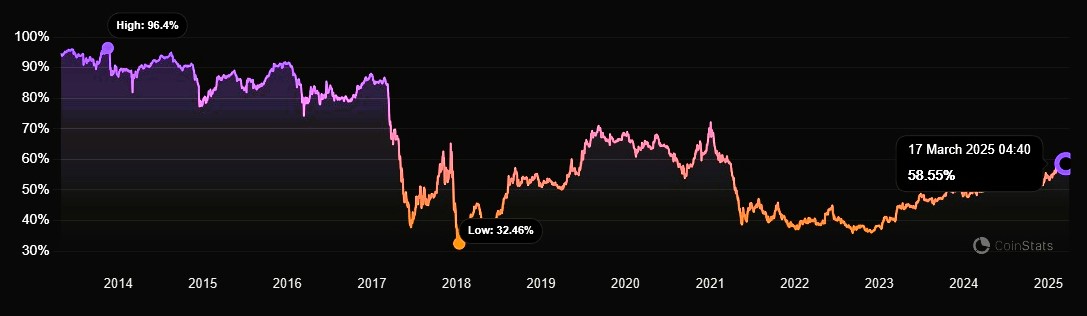}
  \caption{The percentage of market capitalisation of Bitcoin from May 2015 to March 2025.}
  \label{fig:capital_comparison}
\end{figure*}

Despite its relatively short history, the cryptocurrency market has become a significant component of global finance~\cite{gajardo2018does}. While Bitcoin remains the most well-known asset, its dominance is steadily declining. For instance, as shown in Figure~\ref{fig:capital_comparison}, its capital dropped from 85\% in February 2017 to 58\% in March 2025.\footnote{Obtained from coinstats.app in April 2025} Meanwhile, the number of altcoins has grown from around 50 in 2013 to around 10,000 by June 2025,\footnote{www.statista.com} offering innovations that address Bitcoin’s limitations in security, privacy, and stability. This shift underscores the growing importance of analysing altcoins as key players in the cryptocurrency domain, alongside Bitcoin, to better capture the evolving dynamics and investment behaviours in the cryptocurrency market.

As financial assets, cryptocurrencies face challenges similar to those in traditional markets, such as the difficulty of accurate price prediction due to the noisy and nonlinear nature of financial data~\citep{wang2021stock}. However, these challenges are further complicated by the unique characteristics of the cryptocurrency ecosystem. Factors such as mining difficulty, wallet security, social media activity, search trends, interactions with traditional financial assets, and the lack of global acceptance and regulation significantly influence cryptocurrency price movements~\citep{griffith2023cryptocurrency}. Compared to traditional assets, the cryptocurrency market is less mature, and its dynamics are still evolving \cite{amirzadeh2024dynamic}. Therefore, understanding these diverse and interacting drivers is crucial for developing accurate models of cryptocurrency price behaviour.

While it is widely accepted that external factors, such as macroeconomic indicators and social media activity, can influence cryptocurrency prices~\cite{erzurumlu2020one}, most existing studies tend to examine these influences separately. The literature typically focuses either on macroeconomic variables or on the role of sentiment and public discourse~\citep{inamdar2019predicting}. As a result, there is a lack of comprehensive models that jointly evaluate these diverse influences. This separation of factors highlights a key gap in the literature: the need for integrated approaches that assess how financial and social signals collectively shape altcoin price dynamics~\citep{poyser2019exploring}.

In addition, most existing research remains heavily Bitcoin-centric, with studies addressing market efficiency, portfolio strategies, or financial contagion~\citep{kurihara2017market, guesmi2019portfolio, ferreira2019contagion}. Even when altcoins are included, analyses are often generalised rather than coin-specific, overlooking the unique behaviours and drivers of individual assets~\citep{wang2019cryptocurrency}.

Despite growing interest in understanding cryptocurrency price drivers and forecasting, existing modelling approaches often rely on traditional statistical methods or black-box machine learning techniques. These methods typically lack the ability to incorporate domain knowledge, primarily capture correlational rather than causal relationships, and rarely support transparent reasoning about the underlying drivers of price movements. To address these shortcomings, we propose the use of  BNs as a principled framework for modelling the interplay of financial and social factors. BNs identify causal relationships~\citep{sevinc2020bayesian} using directed acyclic graphs (DAGs), and offer key advantages such as handling missing data, integrating expert knowledge with empirical evidence, and enabling explainable inference~\citep{heckerman2008tutorial}, making them a compelling choice for tackling the complexities of cryptocurrency price prediction~\citep{amirzadeh2022applying}.

However, despite these advantages, the application of BNs on cryptocurrency remains limited. Existing studies often neglect the importance of discretisation preprocessing step, applying them without systematic evaluation, regardless of its major impact on BN structure and predictive performance~\cite{nojavan2017comparative}. To address this, we propose a structured discretisation pipeline using multiple methods and bin sizes to identify optimal configurations for cryptocurrency modelling, improving both accuracy and interpretability. Using this pipeline, we construct nine BNs per cryptocurrency considered in this study and evaluate their predictive performance using four different metrics to identify the most accurate models and uncover the key price-driving factors.

Building on these motivations and methodological gaps in the literature, this study makes the following contributions.
\begin{itemize}
    \item Provides a coin-specific modelling framework that uncovers the unique drivers of price movements for each cryptocurrency, allowing tailored analysis and interpretation.
    \item Presents a BN-based modelling framework that jointly incorporates macro-financial indicators and social media signals to capture external influences on cryptocurrency prices
    \item Proposes a structured discretisation pipeline, systematically comparing methods and bin settings to optimise BN construction on continuous time series data.
    \item Enhances interpretability through post-hoc BN analyses, including inference, sensitivity, and influence strength evaluations, offering transparent explanations of price dynamics.
\end{itemize}

The remainder of the paper is structured as follows: Section \ref{Background} reviews the literature on cryptocurrency price factors, while Section \ref{Material and Methods} introduces BNs.  Section~\ref{framwork} outlines the research framework, followed by BN construction in Section~\ref{Experimental}. Section~\ref{Results} presents results, and Section~\ref{Conclusions} concludes with future research directions.


\section{Related literature}  \label{Background}

Research on the relationship between cryptocurrencies and traditional financial assets has predominantly examined correlations, with a focus on identifying diversification opportunities. Corbet et al.~\cite{corbet2018exploring} study the connectedness between cryptocurrencies and traditional financial assets using the concept of spillovers, where an economic event affects seemingly unrelated assets~\cite{lau2017return}. Their findings suggest cryptocurrencies are relatively isolated from financial assets, which is beneficial for diversification opportunities. Charfeddine et al.~\cite{charfeddine2020investigating} examine the dynamic relationships between Bitcoin, Ethereum, and major financial assets like gold, oil, USD/YUAN, and S\&P 500, concluding that correlations are weak and influenced by external shocks. To explore causality, Ji et al.~\cite{ji2018network} use a PC algorithm to construct DAGs between Bitcoin and other financial assets. Their findings indicate that Bitcoin is largely isolated, although time-variant causal relationships appear during bear markets.

Several studies have examined correlations among cryptocurrencies. Stosic et al.~\cite{stosic2018collective} analyse cross-correlations among 119 cryptocurrencies, revealing a complex hierarchical structure that is not apparent when considering only partial correlations. In a similar study, Shi et al.~\cite{shi2020correlations} use a multivariate factor stochastic volatility model and observe smaller groups with similar price volatility among six cryptocurrencies. Bouri et al.~\cite{bouri2021return} study the market integration of 12 cryptocurrencies using a dynamic equicorrelation model. They conclude that trading volume and external uncertainties drive market integration, which varies over time.

Because correlation is important for portfolio optimisation, researchers have explored relationships between cryptocurrencies and traditional financial assets to better understand their price movements. Aslanidis et al.~\cite{aslanidis2019analysis} use a generalised dynamic conditional correlation model to examine relationships between cryptocurrencies, gold, and stock and bond indices, concluding that cryptocurrencies show limited correlation with traditional assets but positive, time-varying correlations among themselves. Similarly, Giudici et al.~\cite{giudici2021crypto} find Bitcoin price is not affected by traditional financial assets, while Malladi et al.~\cite{malladi2021time} demonstrate that different cryptocurrencies react differently to changes in global stock markets and gold. Specifically, they observe that Bitcoin returns are independent of stock market returns but linked to Ripple's return.

Researchers have extended their focus beyond traditional markets, examining additional factors that influence cryptocurrency prices, particularly the role of social media. Lamon et al.~\cite{lamon2017cryptocurrency} use labelled daily news and social media data to determine trends in cryptocurrency prices. Abraham et al.~\cite{abraham2018cryptocurrency} find that the sentiment of the tweet is highly correlated with price movements. Similarly,  Rouhani et al.~\cite{rouhani2019crypto} analyse five million tweets and find that Ripple had the highest percentage of positive tweets (52\%), while Bitcoin received the highest proportion of negative tweets~(27\%). The predictive power of social media sentiment has been further examined by \cite{kraaijeveld2020predictive} and \cite{kim2021dynamics}, showing that Twitter sentiment can predict price returns of Bitcoin, Bitcoin Cash, and Litecoin, with markets more responsive to positive sentiment during downward trends.

In summary, while previous studies have examined both financial correlations and social sentiment, there is still limited understanding of causal structures, particularly in the altcoin domain. Most existing research either focuses on macroeconomic factors or social media influences in isolation. This highlights the need for a more comprehensive approach that combines both dimensions to better explain altcoin price movements.

\section{An overview of Bayesian networks}
\label{Material and Methods}

BNs, a class of probabilistic graphical models, represent complex probability distributions using DAGs~\citep{larranaga2012review}. In a BN, nodes denote random variables and directed edges represent conditional dependencies through DAG, where edges have a defined direction (arrows) and do not form cycles (loops). Based on Bayes’ theorem, BNs support causal inference by factorising the joint distribution, where each node depends only on its direct parents, summarised in a conditional probability table (CPT)~\citep{koller2009probabilistic}.

In BN terminology, a node’s ancestors are nodes from which a directed path reaches it, while descendants are nodes that receive directed paths. Nodes with no descendants are leaf nodes, whereas nodes without ancestors are root nodes \citep{yu2015modified}. The efficiency of BNs stems from their structure, where each variable is conditionally dependent only on its immediate parents and ancestors in the network. For a given set of random variables $\{X_1, X_2, \ldots, X_n\}$ the joint probability distribution $P(X_1, X_2, \ldots, X_n)$ requires $O(2^n)$ parameters for binary variables, which can be computationally intensive. However, by considering only the local dependencies, it can be represented as:
$$P(X_1, X_2, \ldots, X_n)=\prod\limits_{i=1}^n P(X_i|\text{parents}(X_i))$$
where $\text{parents}(X_i)$ is the set of parents of $X_i$ in a DAG representation~\citep{heckerman2008tutorial}.

The nodes in a BN can be continuous or discrete; the discrete nodes are more common. Since most BN learning algorithms operate more efficiently on discrete variables, continuous data are typically discretised using common discretisation methods such as k-means and equal-width binning. The probability of each
discrete variable’s state is conditional on the states of its parents.

A BN can be constructed manually using expert knowledge, automatically using training data, or through a combination of both \citep{zhang2016expert}. Learning a BN structure from data is a challenging task, even with complete data, as it involves structure learning for DAG and parameter estimation for CPTs \citep{de2011efficient, amirkhani2016exploiting}.  These are constructed by calculating probabilities for all possible state combinations of a node and its parents, which can be computationally demanding.

Learning BN structures from data is an NP-hard problem, making it difficult to obtain a complete causal model without considering numerous variables \citep{gheisari2016bnc}.
Two main approaches to structure learning are Bayesian and constraint-based techniques. In the Bayesian approach, an initial structure is built using domain knowledge and then refined with data, whereas constraint-based algorithms search for conditional dependencies to determine relationships \citep{uusitalo2007advantages}.

\section{A Conceptual framework for external drivers of cryptocurrency prices }\label{framwork}

\begin{figure*}[h]
  \centering
\includegraphics[width=0.6\linewidth,height=0.5\textheight,keepaspectratio]
  {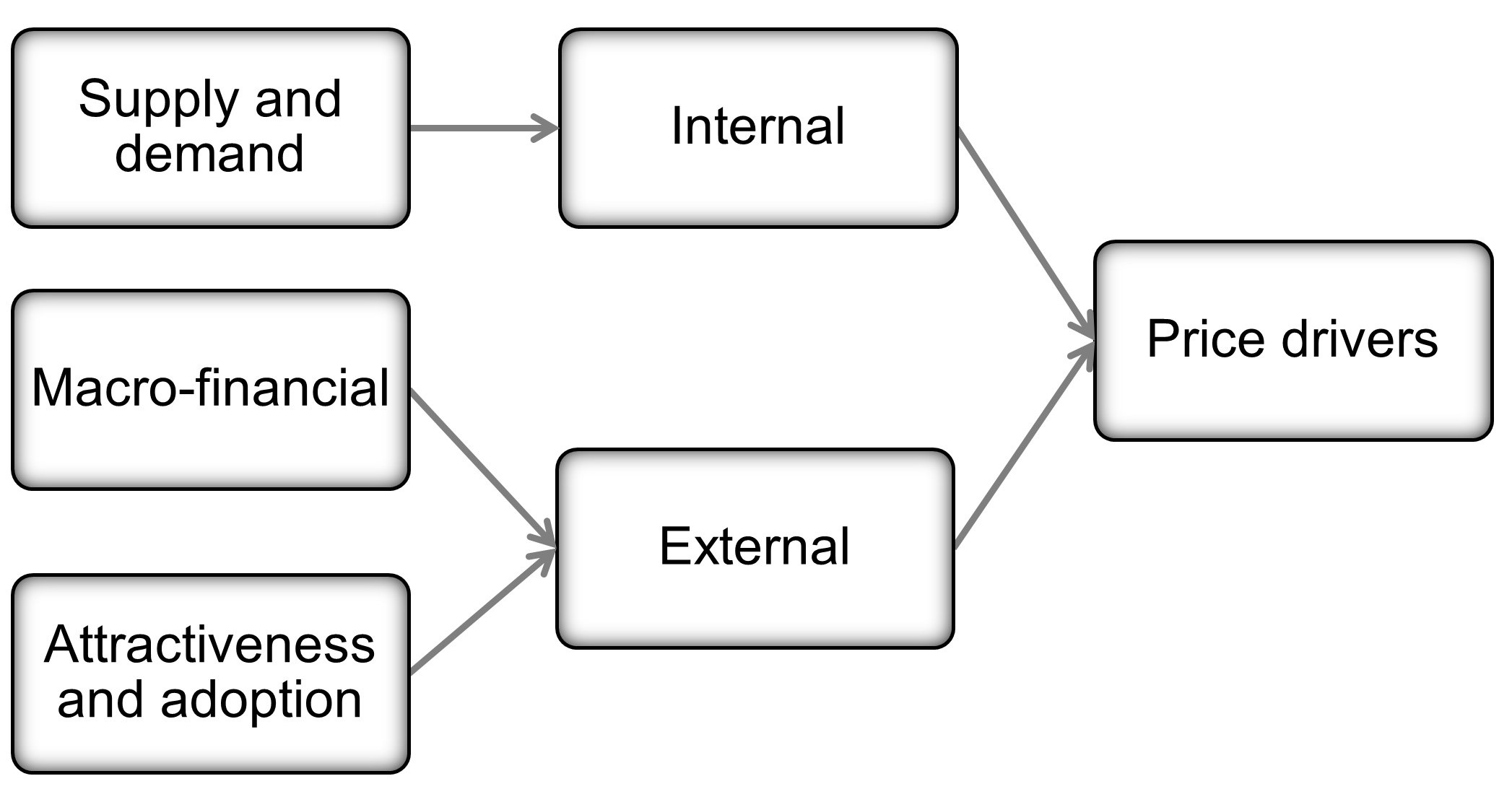}
  \caption{Conceptual framework categorising internal and external factors influencing cryptocurrency prices. Adapted from \cite{poyser2019exploring}.}
  \label{fig:price_drivers}
\end{figure*}

To study the factors driving cryptocurrency prices, we build on the findings of~\cite{ciaian2016economics} for Bitcoin price formation and~\cite{poyser2019exploring} to categorise these factors, as presented in Figure~\ref{fig:price_drivers}. The framework divides price determinants into internal and external categories. As the name implies, internal factors refer to elements that are directly linked to the cryptocurrency ecosystem itself, including supply, demand, transaction volume, and mining difficulty \citep{iccelliouglu2019investigation}. External factors, which are the focus of this study, include macro-financial conditions and investor behaviour—specifically, attractiveness and adoption. In behavioural finance, attractiveness refers to the attractiveness of new financial products, while adoption indicates investors' intention to use these products, such as cryptocurrencies~\citep{ricciardi2000behavioral}. It should be noted that internal and external factors are not completely independent; they dynamically interact in real-world scenarios.

To investigate macro-financial drivers of altcoin prices, we select representative assets from key financial classes: currencies, commodities, and stock indices. We include the S\&P 500 as a proxy for the performance of the US stock market \citep{balcilar2021dynamic}, and two major commodities, gold and West Texas Intermediate (WTI) oil. The WTI oil price is particularly relevant due to its impact on cryptocurrency mining costs \citep{hayes2017cryptocurrency, okorie2020crude}. The USD index (USDX) is also included for its role in altcoin trading, which captures the value of the dollar relative to a basket of major currencies \citep{hasan2022exploring}. Finally, the MSCI World Index serves as a global benchmark for macroeconomic conditions \citep{amirzadeh2023framework}. Thus, our analysis includes two commodities (gold and oil), two equity indices (MSCI and S\&P 500), and one currency index (USDX) as potential macro-financial drivers of altcoin prices.

Behavioural finance, particularly in the context of attractiveness and adoption, examines the psychological drivers of investor decision-making, with an emphasis on emotional influences~\citep{konigstorfer2020applications, ricciardi2000behavioral}. In contrast to traditional finance, which is based on the efficient market hypothesis assumes that asset prices fully reflect all available information, behavioural finance incorporates investor psychology. It accounts for how emotional states influence the investment decision-making process~\citep{rossi2018efficient, malkiel2003efficient}. With the growth of social media, investors are increasingly influenced by shared stories and experiences, shaping perceptions and trading behaviours~ \citep{gurdgiev2020herding, mai2018does, lund2018power}. According to narrative economics,\footnote{Narrative economics, introduced by Nobel winner Robert J. Shiller, studies the dynamics of popular narratives and stories and how people make economic decisions, such as investment in a volatile speculative asset, based on contagious stories~\citep{shiller2017narrative}} such narratives can rapidly gain attraction and significantly shape investment choices and influence financial decision-making~\citep{shiller2017narrative}. Consequently, recent research increasingly focuses on how social media influences financial markets, particularly cryptocurrencies, highlighting the role of investor attention in driving market dynamics \citep{smales2022investor, lin2021investor, zhu2021investor}.

With approximately 330 million monthly active users, X (formerly Twitter) serves as a major platform for public discussions on cryptocurrencies through tweets, hashtags, and live spaces. These conversations offer valuable data for analysing price movements based on public sentiment~\citep{wolk2020advanced}. In this study, we examine the relationship between daily tweet volumes related to altcoins and their intra-day prices as a proxy for investor attractiveness and adoption.

\section{Experimental design}  \label{Experimental}

This study aims to identify the causal relationships among factors influencing cryptocurrency price movements using BNs. As cryptocurrency price data are continuous, most BN software packages require the data to be discretised for model construction~\citep{death2015good}. Although DAGs can be constructed from continuous data using algorithms such as the PC algorithm, these models often lack post-hoc capabilities such as inference and sensitivity analyses.  Hence, selecting appropriate discretisation techniques and bin numbers is essential for accurate modelling~\citep{nojavan2017comparative}.

To address this, we implement a structured procedure that explores combinations of discretisation methods (equal interval, equal quantile, and k-means) and bin numbers (2, 3, and 4), resulting in a total of 54 BN models constructed across six cryptocurrencies. Each model is trained on 80\% of the data and tested on the remaining 20\%. Evaluation is based on three performance metrics: balanced accuracy, F1 score, and area under the ROC curve (AUC). Balanced accuracy is particularly useful for addressing class imbalance in the test sets, providing a more reliable measure of overall model performance than raw accuracy~\cite{guesne2024mind}.
 
\begin{figure}[h]
 \centering
\includegraphics[width=0.6\linewidth,height=0.6\textheight,keepaspectratio]{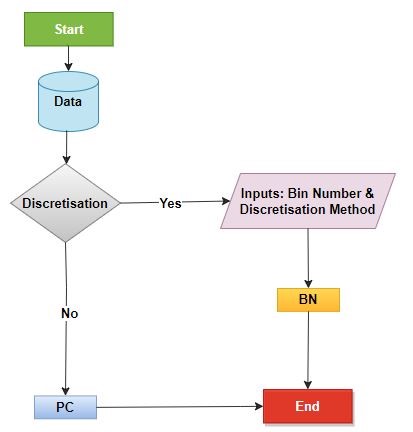}
 \caption{Flowchart illustrating the pipeline for constructing BNs in this study, based on different discretisation methods and bin numbers.}
 \label{fig:alg}
\end{figure}

Since each metric reflects different aspects of model quality, we adopt a composite evaluation strategy to identify the most effective discretisation configuration~\cite{hu2021multisite}. Following the method proposed in~\cite{lee2021neuroimaging}, each metric is scaled to a [0,100] range based on its distribution across all settings using the following formula: 


\begin{equation}
P_{ij} = \left( \frac{S_{ij} - \min V_j}{\max V_j - \min V_j} \right) \times 100, \quad \text{where } V_j = \{ S_{ij} \mid i = 1, 2, \ldots, n \}
\end{equation}
This score provides a unified and interpretable basis for comparing models and is reported as the “Composite Score”.  The best-performing BN for each altcoin is selected for further inference and sensitivity analysis. The overall process is summarised in Figure~\ref{fig:alg}.

\noindent This process involves defining bin-specific state labels and numbers as follows:
\begin{itemize}
 \item \textbf{2-bin}: \textit{Down}, \textit{Up}
\item \textbf{3-bin}: \textit{Down}, \textit{Steady}, \textit{Up}
\item \textbf{4-bin}: \textit{Strong Down}, \textit{Down}, \textit{Up}, \textit{Strong Up}
  \footnote{Increasing the number of bins beyond four also showed no significant improvement in accuracy.}
\end{itemize}

We apply three commonly used  unsupervised ML methods as below~\citep{saleh2018implementation,min2009global}:
\begin{itemize}
  \item \textbf{Equal interval}: Splits the data range into intervals of equal width based on the number of bins. 
  \item \textbf{Equal quantile}: Divides sorted data into equal-sized groups using quantile cutoffs.
  \item \textbf{K-means}: Applies centroid-based clustering to group values into bins, offering flexible partitioning of data~\citep{ahmed2020k, sinaga2020unsupervised}. K-means is particularly noteworthy as it is not commonly included in most BN software packages, yet it often yields improved model performance.
\end{itemize}

Based on this procedure, we construct 54 BNs (6 altcoins × 3 methods × 3 bin counts). Both scaled and raw data are tested, with no significant difference in predictive performance.  Following this setup, all constructed BNs are evaluated using four key performance metrics.



This study employs the GeNIe  software package~\citep{bayesfusion2017genie} to derive BNs. Genie is widely used in the literature for causal modelling and offers essential computational tools.\footnote{Several software packages are available for constructing causal networks, including Netica, Hugin, and Analytica. For a comparative overview, see \cite{mahjoub2011software}.}  We employ the Bayesian Search algorithm with GeNIe’s default configuration. The full set of parameters is listed in Table~\ref{tab:genie_config}. These defaults were selected to ensure consistency, ease of replication, and low computational overhead.
All computations are performed on a system with an Intel Core i7 CPU @ 1.90GHz (2.11 GHz), 16.0 GB RAM, running Windows 10 Enterprise.

\begin{table}[!h]
\centering
\caption{GeNIe parameter settings used for BNs construction.}
\label{tab:genie_config}
\begin{tabular}{ll}
\toprule
\textbf{Parameter}                  & \textbf{Value} \\
\midrule
Learning Algorithm                 & Bayesian Search \\
Discrete Threshold                 & 20 \\
Max Parent Count                   & 8 \\
Iterations                         & 20 \\
Sample Size                        & 50 \\
Seed                               & 0 \\
Link Probability                   & 0.1 \\
Prior Link Probability             & 0.001 \\
\bottomrule
\end{tabular}
\end{table}
 
\subsection{Dataset description and preprocessing}

The cryptocurrencies selected for this study are ranked among the top 10 cryptocurrencies by market capitalisation. Bitcoin, Binance Coin, Ethereum, Litecoin, Ripple, and Tether were selected based on their historical prominence, market capitalisation, and data availability. All selected coins had over 1,200 valid daily observations between January 2018 and April 2023, providing sufficient data length for robust statistical and time series analysis. This extensive period supports reliable correlation assessments with traditional financial, which typically span even longer time horizons. As a result, the selected cryptocurrencies include Bitcoin and five major altcoins, Binance Coin, Ethereum, Litecoin, Ripple, and Tether, which together accounted for approximately 70\% of the total cryptocurrency market capitalisation in 2023.\footnote{Based on the data on coingecko.com}

Daily closing price data for the selected altcoins and financial assets, including gold, MSCI, S\&P 500, WTI, and USDX, were obtained from \texttt{Yahoo Finance} using the \texttt{yfinance} Python library. Additionally, daily tweet volumes related to the altcoins were collected from \texttt{bitinfocharts.com}. Due to the absence of tweet data for Tether, it was excluded from the social media analysis. Summary statistics for both price and tweet data are presented in Appendix~\ref{Prices_Tweet_stats}.

To construct BNs for each cryptocurrency, we first create a merged dataset combining tweet volume, macro-financial indicators, and the corresponding coin’s price data, aligned by date availability. To address the challenges of time series data, we apply several preprocessing steps. This includes handling non-stationarity and distributional irregularities. Extreme outliers are removed using the interquartile range (IQR) method, while stationarity is tested using the Augmented Dickey-Fuller (ADF) and KPSS tests. When necessary, first-order differencing is applied to enforce stationarity. These preprocessing steps, detailed in Appendix~\ref{adf} and Appendix~\ref{Outlier}, are essential for ensuring reliable structure learning and inference in the BN models.

Furthermore, to reduce skewness and improve the consistency of binning during discretisation, we applied a logarithmic transformation to all price variables. This stabilises variance, mitigates extreme values, and supports more balanced binning across discretisation strategies~\citep{tessema2021enhancing}. Summary statistics of the raw data distributions are provided in Appendix~\ref{Outlier}, highlighting the skewness and outlier patterns that motivated this transformation.

\section{Results and discussion} \label{Results}

This section presents the results of identifying the key factors influencing cryptocurrency price movements using BNs, along with post-hoc analyses including inference and sensitivity analysis. Section~\ref{pred_acc} compares the predictive performance of the 54 trained BNs using four metrics. Section~\ref{st_learn} provides detailed results from inference,  sensitivity and the strength of influence analyses. Inference analysis highlights the most influential predictors of the target variable (cryptocurrency prices), while sensitivity analysis quantifies how variations in these predictors affect the probability distribution of the target node. Additionally, the strength of influence analysis measures the magnitude of causal effects between variables, offering further insights into the underlying price dynamics.

\subsection{Predictive performance of Bayesian networks across discretisation settings
} \label{pred_acc}

We evaluate the predictive performance of 54 BN models. Each model predicts directional price movements based on the posterior probabilities of the target node. Performance is assessed using balanced accuracy, F1 Score,  AUC, and Composite Score as reported in Table~\ref{tab:BN_prediction_metrics}. The configuration with the highest composite score for each cryptocurrency is highlighted in bold. The final row shows how often each discretisation–bin combination achieved the top score across all cryptocurrencies.

\begin{table*}[h]
\caption{Prediction performance of BNs using different discretisation strategies and bin settings, evaluated by balanced accuracy, AUC, F1-score, and a Composite Score. The highest value in each row is highlighted.}
\label{tab:BN_prediction_metrics}
\begin{adjustbox}{max width=\textwidth}
\renewcommand{\arraystretch}{1.1}
{\large
\begin{tabular}{|>{\raggedright\arraybackslash}p{2.5cm} >{\raggedright\arraybackslash}p{3.5cm} 
|>{\centering\arraybackslash}p{1.5cm} >{\centering\arraybackslash}p{1.5cm} >{\centering\arraybackslash}p{1.5cm} 
|>{\centering\arraybackslash}p{1.5cm} >{\centering\arraybackslash}p{1.5cm} >{\centering\arraybackslash}p{1.5cm} 
|>{\centering\arraybackslash}p{1.5cm} >{\centering\arraybackslash}p{1.5cm} >{\centering\arraybackslash}p{1.5cm}|}

\hline
\multicolumn{2}{|c|}{\multirow{2}{*}{\textbf{Coin / Metric}}} & \multicolumn{3}{c|}{\textbf{Equal Interval}} & \multicolumn{3}{c|}{\textbf{Equal Quantile}} & \multicolumn{3}{c|}{\textbf{K-means}} \\
\multicolumn{2}{|c|}{} & 2 bins & 3 bins & 4 bins & 2 bins & 3 bins & 4 bins & 2 bins & 3 bins & 4 bins \\
\hline
\multirow{4}{*}{\rotatebox{35}{\textbf{Bitcoin}}} 
& Balanced Acc. & {0.5000} & 0.3333 & 0.2357 & {0.5000} & 0.4938 & 0.2559 & 0.4917 & 0.3471 & 0.2630 \\
& AUC           & 0.5948 & {0.7825} & 0.4791 & 0.5000 & 0.7362 & 0.5227 & 0.4917 & 0.4909 & 0.5847 \\
& F1            & 0.0112 & 0.6442 & 0.5975 & 0.3312 & 0.4941 & 0.2157 & {0.7178} & 0.5807 & 0.1973 \\
& Composite Score & 138.13 & 226.51 & 82.97 & 152.18 & {\large\textit{\textbf{250.74}}} & 50.95 & 201.01 & 126.64 & 71.47 \\ \hline

\multirow{4}{*}{\rotatebox{35}{\textbf{Binance Coin}}} 
& Balanced Acc. & {0.6971} & 0.3241 & 0.2565 & 0.5086 & 0.2849 & 0.2500 & 0.4999 & 0.3333 & 0.2624 \\
& AUC           & 0.6438 & 0.6777 & 0.5423 & 0.5086 & 0.6262 & 0.5443 & 0.3635 & 0.4891 & 0.4334 \\
& F1            & 0.5454 & 0.2307 & 0.3923 & 0.4546 & 0.2631 & 0.0993 & {0.7098} & 0.0005 & 0.3614 \\
& Composite Score & {\large\textit{\textbf{266.03}}} & 149.03 & 113.60 & 168.04 & 128.44 & 71.47 & 155.89 & 58.61 & 75.90 \\ \hline
\multirow{4}{*}{\rotatebox{35}{\textbf{Ethereum}}} 
& Balanced Acc. & 0.4880 & 0.3317 & 0.2908 & {0.5000} & 0.4096 & 0.3438 & 0.4542 & 0.3510 & 0.2500 \\
& AUC           & 0.5371 & 0.1789 & 0.5871 & 0.5000 & {0.6120} & 0.4474 & 0.4429 & 0.4327 & 0.4609 \\
& F1            & 0.7134 & {0.7817} & 0.5784 & 0.3333 & 0.4047 & 0.3410 & 0.2805 & 0.4445 & 0.3731 \\
& Composite Score & {\large\textit{\textbf{246.22}}} & 107.25 & 206.63 & 170.93 & 215.63 & 132.44 & 166.15 & 139.82 & 121.80 \\ \hline

\multirow{4}{*}{\rotatebox{35}{\textbf{Litecoin}}}
& Balanced Acc. & {0.5000} & 0.3338 & 0.2639 & {0.5000} & 0.4804 & 0.3047 & {0.5000} & 0.3273 & 0.2723 \\
& AUC           & 0.6363 & 0.1789 & 0.5601 & 0.4531 & {0.7287} & 0.4583 & 0.6578 & 0.4857 & 0.4424 \\
& F1            & {0.8560} & 0.3787 & 0.6281 & 0.3333 & 0.4804 & 0.2500 & 0.4954 & 0.4661 & 0.4001 \\
& Composite Score & {\large\textit{\textbf{{283.19}}}} & 50.84 & 131.73 & 163.62 & 229.72 & 68.10 & 227.60 & 118.32 & 76.25 \\ \hline

\multirow{4}{*}{\rotatebox{35}{\textbf{Ripple}}}
& Balanced Acc. & 0.5000 & 0.3318 & 0.2518 & {0.5195} & 0.3362 & 0.2266 & 0.5000 & 0.3333 & 0.2718 \\
& AUC           & 0.6123 & 0.2371 & 0.4380 & 0.5195 & {0.6181} & 0.4638 & 0.4814 & 0.4982 & 0.4453 \\
& F1            & 0.6481 & {0.8493} & 0.6357 & 0.5133 & 0.3045 & 0.1927 & 0.2120 & 0.0187 & 0.3364 \\
& Composite Score & {\large\textit{\textbf{267.60}}} & 135.92 & 135.62 & 233.67 & 171.83 & 80.45 & 180.74 & 104.96 & 108.33 \\ \hline

\multirow{4}{*}{\rotatebox{35}{\textbf{Tether}}}
& Balanced Acc. & {0.5000} & 0.3333 & 0.2405 & {0.5000} & 0.3333 & 0.2500 & {0.5000} & 0.3333 & 0.2500 \\
& AUC           & {0.5000} & {0.5000} & 0.4778 & {0.5000} & {0.5000} & {0.5000} & {0.5000} & {0.5000} & {0.5000} \\
& F1            & 0.1930 & {0.9827} & 0.2359 & 0.3333 & 0.1689 & 0.0986 & {0.9886} & 0.0001 & 0.0000 \\
& Composite Score & 98.94 & 260.83 & 97.31 & 164.55 & 88.33 & 72.84 & {\large\textit{\textbf{261.75}}} & 50.00 & 50.00 \\ \hline
\multirow{1}{*}{\textbf{Number of winning}}                 &   & 4                & 0           & 0                & 0           &1                & 0           & 1                & 0           & 0           \\ \hline
\end{tabular}
}
\end{adjustbox}
\end{table*}

Based on the composite score, the equal interval method with two bins performs best, achieving the top score in four out of six cryptocurrencies—more than any other configuration. K-means and equal quantile with two or three bins show competitive but less consistent results, each producing a top score only once. These findings suggest that simpler discretisation strategies, especially Equal intervals with two bins, offer more reliable performance by maintaining class separability and avoiding over-fragmentation.

\begin{figure}[!h]
  \centering
\includegraphics[width=0.8\linewidth,height=0.4\textheight,keepaspectratio]{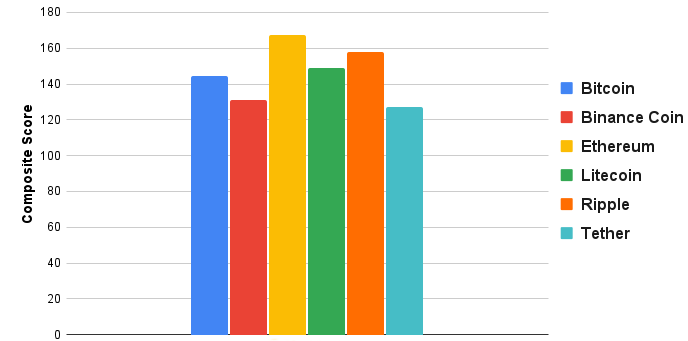}
  \caption{Average Composite Score for each cryptocurrency, aggregated across all discretisation strategies and bin configurations, based on the results reported in Table~\ref{tab:BN_prediction_metrics}.}
  \label{fig:avg_composite_score}
\end{figure}

To compare overall model performance across cryptocurrencies, we calculate the average composite score for each coin by aggregating results across all discretisation methods and bin settings. As shown in Figure~\ref{fig:avg_composite_score}, Ethereum achieves the highest average score~(167.33), followed by Ripple and Litecoin, whereas Tether and Binance Coin yield the lowest. These results suggest that certain cryptocurrencies possess data structures that better align with the probabilistic assumptions of BNs.  

\subsection{Discretisation impact on cryptocurrency}
\label{sec:coin_discretisation}

\begin{figure*}[h]
  \centering
\includegraphics[width=\linewidth]{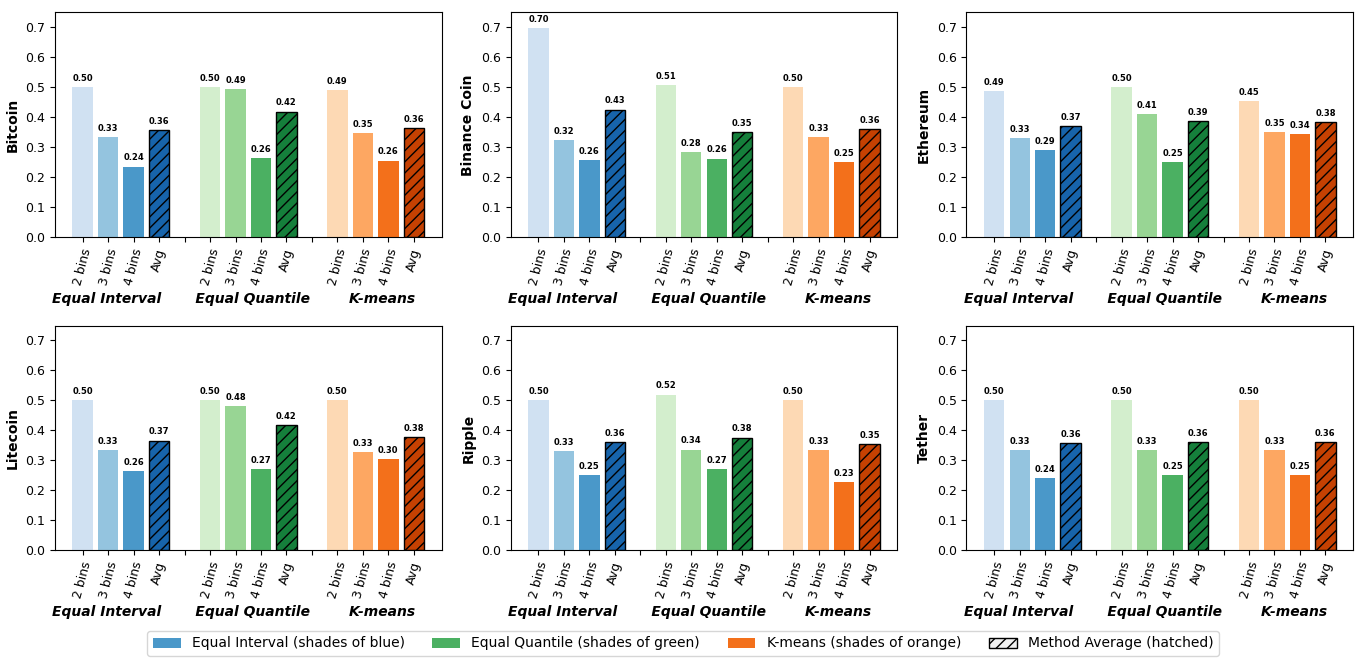}
 \caption{Balanced accuracy scores of all discretisation configurations across cryptocurrencies. Bars represent scores for two, three, and four bins. Hatched bars indicate the average performance for each method.}  \label{fig:coin_specific_scores}
\end{figure*}

This subsection provides a more detailed evaluation of discretisation performance at the individual cryptocurrency level. Based on model performance reported in Table~\ref{tab:BN_prediction_metrics}, Figure~\ref{fig:coin_specific_scores} illustrates results using balanced accuracy, which is selected due to its robustness to class imbalance and suitability for fair comparison. A hatched bar in each group marks the method’s average. This enables comparison across both bin settings and overall discretisation strategies.

Equal interval with two bins yields the highest balanced accuracy for most coins, including Bitcoin, Binance Coin, Ripple, suggesting that simpler binning preserves signal better.
Notably, a general decline in accuracy is observed as the number of bins increases, indicating that finer granularity may introduce noise.
K-means shows greater variability and is more sensitive to bin count.  The hatched bars reinforce the robustness of equal interval across multiple cryptocurrencies.

At the individual coin level, equal interval with two bins yields the best performance for most cases, with Binance Coin achieving the highest balanced accuracy of 0.70. In contrast, the lowest scores often occur with four-bin configurations, especially under K-means for Ripple and Ethereum. While some coins such as Litecoin and Tether show more balanced results across methods, others exhibit clear preferences, highlighting method sensitivity at the coin level.

Increasing the number of bins generally reduces balanced accuracy, with the sharpest declines seen in K-means and equal interval. For example, Binance Coin drops from 0.6971 (Interval–2 bins) to 0.2565 (Interval–4 bins), and Ripple falls from 0.5195 (Quantile–2 bins) to 0.2266 (K-means–4 bins). Equal quantile shows more stable performance across bin sizes but does not reach the top scores. These results highlight the importance of carefully balancing bin granularity with model stability.

\subsection{Evaluating the effect of discretisation on Bayesian networks  performance}

To better understand how discretisation impacts model performance, we first examine the distribution of data points across different binning strategies and then evaluate their predictive outcomes. Table~\ref{tab:Comparison bin} presents the number of data points in each bin for various discretisation strategies and bin settings.

\begin{table*}[t]
\caption{The number of data points in each bin based on the number of intervals and the unsupervised ML method for discretisation. The bottom rows report average bin counts across all cryptocurrencies for each bin configuration.}
\label{tab:Comparison bin}
\begin{adjustbox}{max width=\textwidth}
\begin{tabular}{|c|c|rr|rrr|rrrr|}
\hline
\multirow{2}{*}{\textbf{Cryptocurrency}} & \multirow{2}{*}{\textbf{Method}} &
\multicolumn{2}{c|}{\textbf{2 Intervals}} &
\multicolumn{3}{c|}{\textbf{3 Intervals}} &
\multicolumn{4}{c|}{\textbf{4 Intervals}} \\
& & \multicolumn{1}{c|}{\textit{Bin1}} & \textit{Bin2} &
\multicolumn{1}{c|}{\textit{Bin1}} & \multicolumn{1}{c|}{\textit{Bin2}} & \textit{Bin3} &
\multicolumn{1}{c|}{\textit{Bin1}} & \multicolumn{1}{c|}{\textit{Bin2}} & \multicolumn{1}{c|}{\textit{Bin3}} & \textit{Bin4} \\ 
\hline

\multirow{3}{*}{\textbf{Bitcoin}} & EqualInterval & 20 & 237 & 3 & 193 & 61 & 2 & 18 & 223 & 14 \\
 & EqualQuantile & 128 & 129 & 86 & 85 & 86 & 64 & 64 & 64 & 65 \\
 & K-means & 225 & 32 & 176 & 64 & 17 & 179 & 46 & 30 & 2 \\ \arrayrulecolor{gray}\hline
\multirow{3}{*}{\textbf{Binance Coin}} & EqualInterval & 12 & 241 & 4 & 100 & 149 & 3 & 9 & 190 & 51 \\ 
 & EqualQuantile & 126 & 127 & 85 & 84 & 84 & 63 & 63 & 63 & 64 \\
 & K-means & 51 & 202 & 96 & 4 & 153 & 150 & 42 & 57 & 4 \\  \arrayrulecolor{gray}\hline
\multirow{3}{*}{\textbf{Ethereum}} & EqualInterval & 44 & 212 & 9 & 216 & 31 & 3 & 41 & 201 & 11 \\
 & EqualQuantile & 128 & 128 & 86 & 85 & 85 & 64 & 64 & 64 & 64 \\
 & K-means & 41 & 215 & 147 & 87 & 22 & 85 & 137 & 14 & 20 \\  \arrayrulecolor{gray}\hline
\multirow{3}{*}{\textbf{Litecoin}} & EqualInterval & 25 & 231 & 6 & 137 & 113 & 3 & 22 & 184 & 47 \\
 & EqualQuantile & 128 & 128 & 86 & 85 & 85 & 64 & 64 & 64 & 64 \\
 & K-means & 163 & 93 & 155 & 74 & 27 & 135 & 48 & 66 & 7 \\  \arrayrulecolor{gray}\hline
\multirow{3}{*}{\textbf{Ripple}} & EqualInterval & 63 & 193 & 10 & 224 & 22 & 4 & 59 & 184 & 9 \\
 & EqualQuantile & 128 & 128 & 86 & 85 & 85 & 64 & 64 & 64 & 64 \\
 & K-means & 98 & 158 & 26 & 31 & 199 & 135 & 10 & 21 & 90 \\  \arrayrulecolor{gray}\hline
\multirow{3}{*}{\textbf{Tether}} & EqualInterval & 167 & 95 & 2 & 259 & 1 & 1 & 166 & 94 & 1 \\
 & EqualQuantile & 131 & 131 & 88 & 86 & 88 & 66 & 65 & 65 & 66 \\
 & K-means & 260 & 2 & 259 & 2 & 1 & 51 & 2 & 1 & 208 \\
 \arrayrulecolor{black}\hline
\multirow{3}{*}{\textbf{Average}} & {EqualInterval} & 55.2 & 201.5 & 5.7 & 188.2 & 62.8 & 2.7 & 52.5 & 179.3 & 22.2 \\
                                  & {EqualQuantile} & 128.2 & 128.5 & 86.2 & 85.0 & 85.5 & 64.2 & 64.0 & 64.0 & 64.5 \\
                                  & {K-means}       & 139.7 & 117.0 & 143.2 & 43.7 & 69.8 & 122.5 & 47.5 & 31.5 & 55.2 \\
\hline

\end{tabular}
\end{adjustbox}
\end{table*}

Equal interval tends to produce skewed distributions, particularly toward the \textit{Up} class in the 2-bin setting (201.5 vs 55.2) and the \textit{Steady} class in the 3-bin setting (188.2 on average), which shows its sensitivity to how the data is spread numerically. As expected from its design, the equal quantile method yields nearly uniform distributions across bins, with the 2-bin configuration averaging 128.2 and 128.5 for \textit{Down} and \textit{Up}, respectively. The 3-bin and 4-bin settings also show similarly balanced splits. 

K-means discretisation reflects the underlying cluster structure but often results in significant class imbalance. For instance, the 3-bin configuration assigns an average of 143.2 instances to \textit{Down} compared to 43.7 to \textit{Steady}. These results suggest that while K-means may capture some hidden patterns in the data, it can also create dominant classes that reduce model flexibility.

Furthermore, Table~\ref{tab:Comparison bin} also reveals notable differences in bin balance across cryptocurrencies. Tether consistently shows the most imbalanced distributions, particularly under K-means and equal interval methods, where one bin dominates (e.g., 260 vs 2 in the 2-bin K-means case). In contrast, Bitcoin and Ethereum tend to exhibit more balanced distributions—especially under the equal quantile method, which yields near-uniform splits across all bin counts. This suggests that some coins are better suited to balanced discretisation due to their underlying distribution.

\subsection{Bayesian networks reasoning for cryptocurrency price movements  
}  \label{st_learn}

This section presents inference and sensitivity analyses for the top-performing BNs identified for each cryptocurrency based on the highest composite score, following methodologies used by \cite{asvija2021security, hosseini2019development}.   Each selected BN is visualised alongside its inference and sensitivity analysis outputs in Figures~\ref{fig:Inference} and~\ref{fig:sensitivity}, respectively.

\subsubsection{Inference analysis}

\begin{figure*}[h]
\centering
\begin{minipage}{\textwidth}
\centering

\begin{subfigure}[t]{0.48\textwidth}
  \centering
  \fbox{\includegraphics[width=\linewidth, height=0.12\textheight]{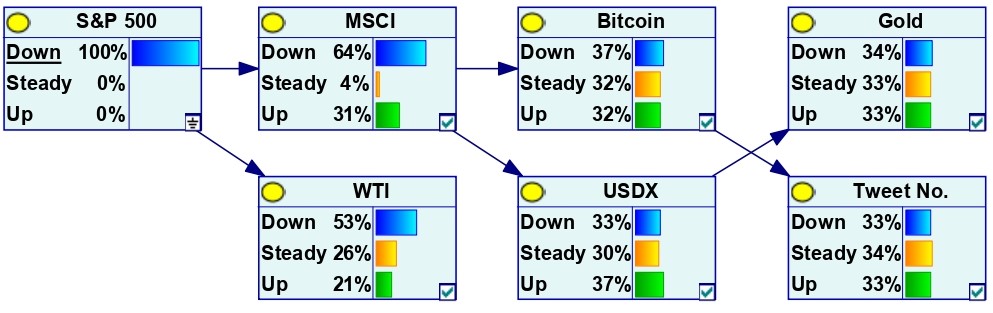}}
 \caption{Bitcoin (Equal quantile, 3 bins)}

  \label{fig:BTC_inf}
\end{subfigure}
\hfill
\begin{subfigure}[t]{0.48\textwidth}
  \centering
   \fbox{\includegraphics[width=\linewidth, height=0.12\textheight]{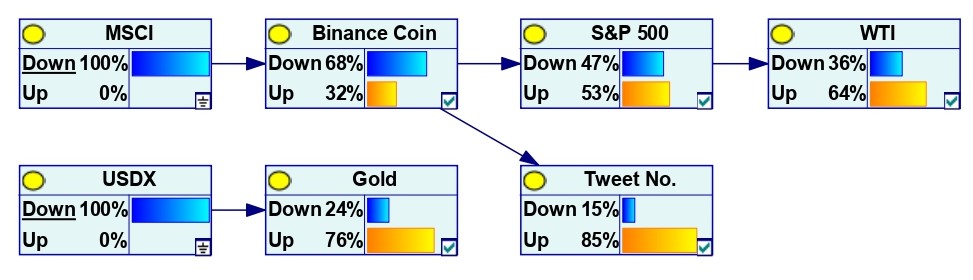}}
  \caption{Binance Coin (Equal interval, 2 bins)}
  \label{fig:BNB_inf}
\end{subfigure}

\vspace{0.2cm}

\begin{subfigure}[t]{0.48\textwidth}
  \centering
  \fbox{\includegraphics[width=\linewidth, height=0.12\textheight]{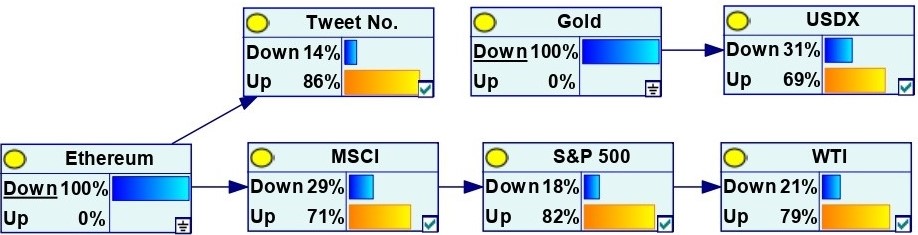}}
  \caption{Ethereum (Equal interval, 2 bins)}
  \label{fig:ETH_inf}
\end{subfigure}
\hfill
\begin{subfigure}[t]{0.48\textwidth}
  \centering
  \fbox{\includegraphics[width=\linewidth, height=0.12\textheight]{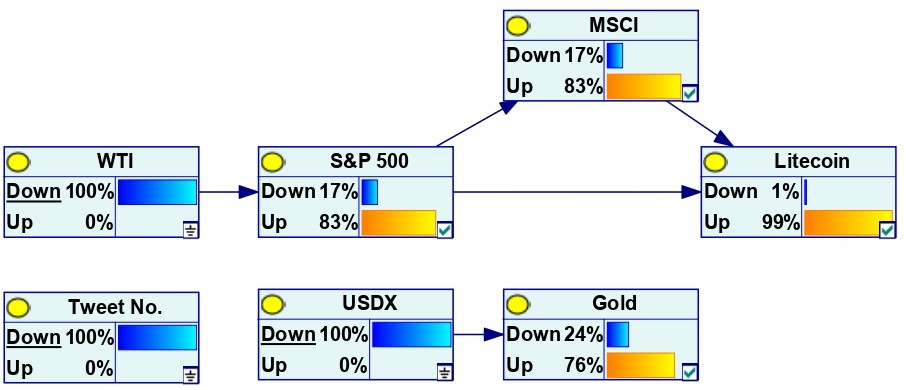}}
 \caption{Litecoin (Equal interval, 2 bins)}
  \label{fig:LTC_inf}
\end{subfigure}

\vspace{0.2cm}

\begin{subfigure}[t]{0.48\textwidth}
  \centering
  \fbox{\includegraphics[width=\linewidth, height=0.12\textheight]{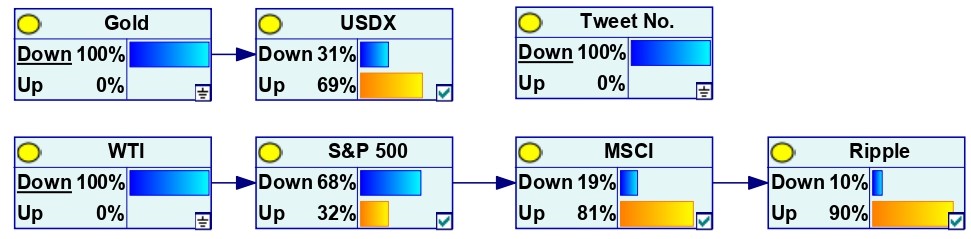}}
 \caption{Ripple (Equal interval, 2 bins)}
  \label{fig:XRP_inf}
\end{subfigure}
\hfill
\begin{subfigure}[t]{0.48\textwidth}
  \centering
  \fbox{\includegraphics[width=\linewidth, height=0.12\textheight]{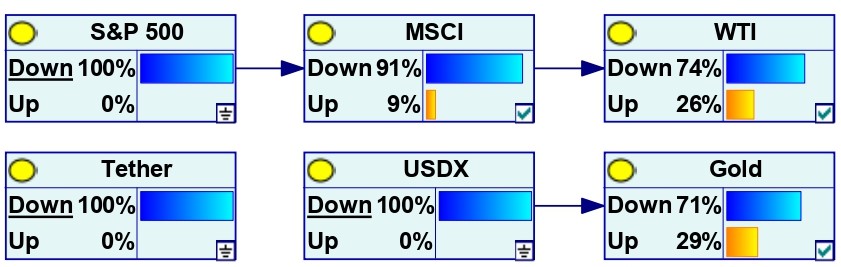}}
 \caption{Tether (K-means, 2 bins)}
  \label{fig:Tether_inf}
\end{subfigure}

\caption{Inference diagrams for the best-performing BNs of each cryptocurrency. In each network, the root node representing the target variable (price direction) is fixed to the `Down' state, and the resulting posterior beliefs of connected variables are observed. This illustrates how evidence propagates through the network via probabilistic reasoning.}

\label{fig:Inference}
\end{minipage}
\end{figure*}

 Inference in BNs is a form of probabilistic reasoning that examines how changes in one variable propagate through the network. Specifically, we observe changes in  CPTs of other nodes when a particular node is fixed in one of its states. Figure~\ref{fig:Inference} presents inference analysis for the top-performing BNs. In our analysis, each cryptocurrency is set as the target node to illustrate how its associated factors respond. Inference can be bi-directional, used for prediction (cause to effect) or diagnosis (effect to cause)~\citep{lu2020risk}.

From the perspective of individual cryptocurrencies, BNs reveal various levels of connectivity. Using graph terminology,\footnote{The number of directed edges entering a node $v$, known as the in-degree $\text{deg}^-(v)$, quantifies how many variables directly influence that node. Similarly, the out-degree $\text{deg}^+(v)$ represents how many variables the node influences. The total degree is given by $\text{deg}(v) = \text{deg}^-(v) + \text{deg}^+(v)$, reflecting the overall connectivity of the node within the graph.} based on Figure~\ref{fig:BTC_inf}, Bitcoin has an in-degree of $\text{deg}^-(\text{Bitcoin}) = 1$. This dependency suggests that global equity changes~(MSCI) may play a key role in shaping Bitcoin's directional movements. Notably, Bitcoin also acts as a parent node, exerting influence on Tweet No., implying that its price behaviour may drive changes in social media activity and sentiment.

As depicted in Figure~\ref{fig:BNB_inf}, $\text{deg}-(\text{Binance\ Coin}) = 3$, with edges connecting to MSCI, S\&P 500, and Tweet No. Similar to Bitcoin, Binance Coin has a direct incoming edge from MSCI. Additionally, the presence of an edge from S\&P 500 indicates a broader financial sensitivity compared to Bitcoin. This combination suggests that Binance Coin's directional movements are more strongly influenced by traditional financial assets. Moreover, Binance Coin also influences Tweet No., implying that its price behaviour may drive social media sentiment, as observed in the Bitcoin network.

Unlike other networks, Ethereum~(Figure~\ref{fig:ETH_inf}) appears structurally more isolated from financial assets, with only a single incoming edge to MSCI. Notably, Gold and USDX are completely disconnected and do not influence Ethereum's price movements. Moreover, Ethereum is also a parent of Tweet No., similar to Bitcoin and Binance Coin, indicating that its directional movements may play a dominant role in shaping social sentiment.

As shown in Figure~\ref{fig:LTC_inf}, Litecoin receives direct inputs from both MSCI and S\&P 500, resulting in an in-degree of $\text{deg}^-(\text{Litecoin}) = 2$. Considering the network state is set to \texttt{Down}, this reflects a relatively high level of connectivity. Litecoin remains well integrated with these financial assets and shows a highly confident posterior prediction, with a 99\% probability for an upward movement. Notably, Litecoin is isolated from several variables in the network, including USDX, WTI, and Tweet No.

As shown in Figure~\ref{fig:XRP_inf}, $\text{deg}^-(\text{Ripple}) = 1$, with a direct edge from MSCI. This structure reflects a linked path from commodity prices through equity markets to Ripple’s price direction. Notably, Ripple is completely disconnected from social sentiment (Tweet No.) as well as from Gold and USDX under this configuration.

Tether~(Figure~\ref{fig:Tether_inf}) has no incoming edges, where $\text{deg}^-(\text{Tether}) = 0$, and it is completely isolated in this network, with no edge influence from any financial. As a stablecoin, this may reflect its distinct role within the cryptocurrency ecosystem, unlike mining-based assets, Tether is typically pegged to fiat currency and designed to maintain price stability, which may reduce its sensitivity to broader market signals under this configuration.

Across the networks, a consistent pattern emerges regarding social sentiment. For Bitcoin, Ethereum, and Binance Coin, the price direction nodes act as parents of Tweet No., indicating that their price movements may influence tweet volume or sentiment, rather than being influenced by it. This suggests that social media activity tends to lag behind market movements, reflecting public reactions to price changes rather than driving them. In contrast, Litecoin, Ripple, and Tether have no edge to Tweet No., suggesting that their price movements may not be directly influenced by social media activity.

Considering the financial assets, MSCI appears as a parent node in four of the BNs, indicating its significant role as a causal factor in cryptocurrency price movements. This highlights its importance as a global equity indicator. S\&P 500 also frequently has direct edges to cryptocurrencies, reinforcing the relevance of traditional equity markets in shaping crypto price dynamics. In contrast, Gold and USDX often appear as isolated nodes across several BNs, suggesting their limited influence.

\subsubsection{Sensitivity and influence strength}

Sensitivity analysis in BNs examines how variations in input factors influence changes in the target variable~\citep{castillo1997sensitivity}. It quantifies the extent to which uncertainty in the model's output stems from uncertainty in its inputs~\citep{saltelli2019so}, offering insights into causal relationships and the model’s internal logic~\citep{meurisse2022risk}. As a common tool in economic evaluations, it helps analysts to assess the reliability of outcomes , and aids in evaluating decision reliability~\citep{walker2001allowing}. In this study, we use sensitivity analysis to identify and prioritise key price-driving factors for each altcoin, guiding investment strategies and portfolio management.



\begin{figure*}[h]
\centering
\begin{minipage}{\textwidth}
\centering

\begin{subfigure}[t]{0.48\textwidth}
  \centering
 \fbox{\includegraphics[width=\linewidth, height=0.12\textheight]{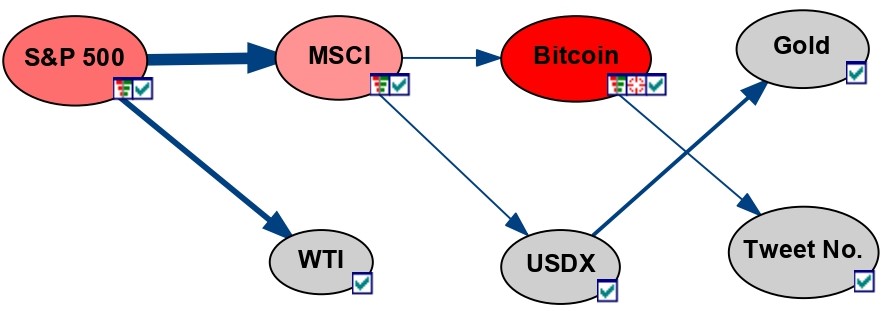}}
 \caption{Bitcoin}

  \label{fig:BTC_sen}
\end{subfigure}
\hfill
\begin{subfigure}[t]{0.48\textwidth}
  \centering
  \fbox{\includegraphics[width=\linewidth, height=0.12\textheight]{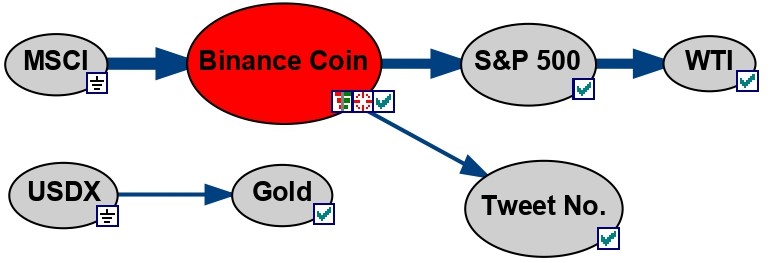}}
  \caption{Binance Coin }
  \label{fig:BNB_sen}
\end{subfigure}

\vspace{0.2cm}

\begin{subfigure}[t]{0.48\textwidth}
  \centering
\fbox{\includegraphics[width=\linewidth, height=0.12\textheight]{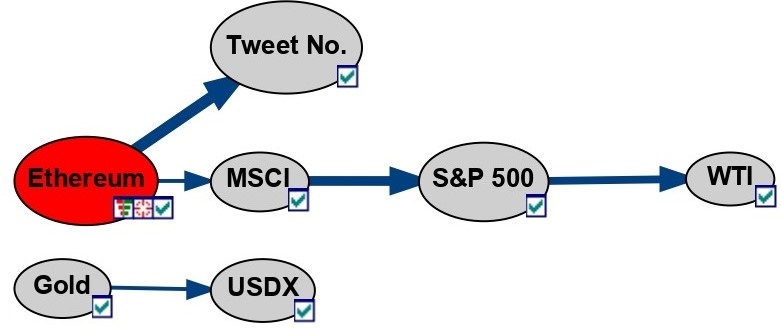}}
  \caption{Ethereum }
  \label{fig:ETH_sen}
\end{subfigure}
\hfill
\begin{subfigure}[t]{0.48\textwidth}
  \centering
 \fbox{\includegraphics[width=\linewidth, height=0.12\textheight]{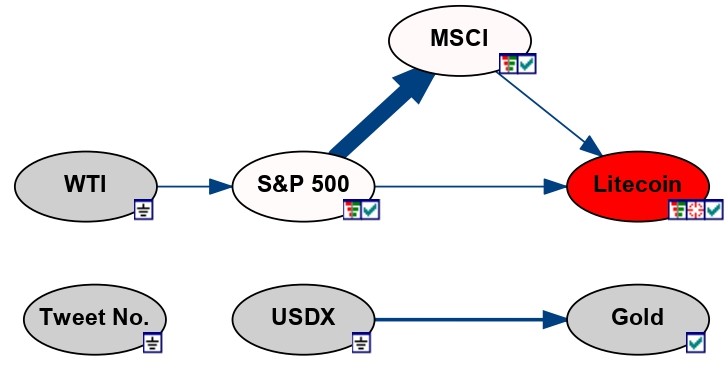}}
 \caption{Litecoin}
  \label{fig:LTC_sen}
\end{subfigure}

\vspace{0.2cm}

\begin{subfigure}[t]{0.48\textwidth}
  \centering
  \fbox{\includegraphics[width=\linewidth, height=0.12\textheight]{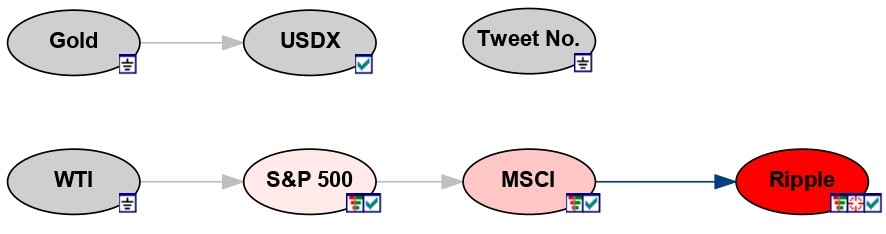}}
 \caption{Ripple}
  \label{fig:XRP_sen}
\end{subfigure}
\hfill
\begin{subfigure}[t]{0.48\textwidth}
  \centering
 \fbox{\includegraphics[width=\linewidth, height=0.12\textheight]{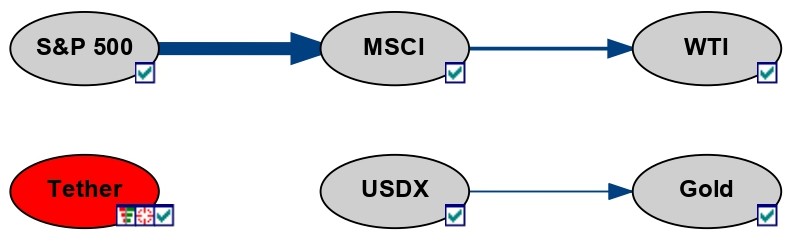}}
 \caption{Tether}
  \label{fig:Tether_sen}
\end{subfigure}
\caption{Sensitivity and strength-of-influence analysis for each cryptocurrency. Each altcoin is set as the target node, and red shading indicates the relative sensitivity of input variables. Arc thickness represents the strength of influence between connected variables, with thicker lines denoting stronger causal relationships.}

\label{fig:sensitivity}
\end{minipage}
\end{figure*}

The sensitivity analysis was conducted using Genie, implementing the algorithm described in \cite{coupe2000computational}. For each cryptocurrency, the corresponding node was set as the target variable within its best-performing BN. The resulting visual outputs highlight the influence of each input factor, with the intensity of red shading representing the relative strength of impact. These visualisations are presented in Figure~\ref{fig:sensitivity} for all six cryptocurrencies.

According to the results of the sensitivity analysis (Figure~\ref{fig:BTC_sen}), Bitcoin is primarily influenced by S\&P 500 and MSCI, with S\&P 500 having the strongest effect. In contrast, Binance Coin, Ethereum, and Litecoin exhibit low sensitivity to other variables, as reflected by the minimal presence of red shading in their diagrams (Figure~\ref{fig:BNB_sen}, \ref{fig:ETH_sen}, and \ref{fig:LTC_sen}).

For Ripple (Figure~\ref{fig:XRP_sen}), MSCI emerges as the dominant influencer, with S\&P 500 contributing more weakly, as can be seen from the lighter colour intensity. This suggests that Ripple’s direction is more closely tied to global equity performance than commodity-driven movements. Furthemore, Tether (Figure~\ref{fig:Tether_sen}) remains fully insensitive to external variables, consistent with its structural isolation observed in the BN inference.

While sensitivity analysis highlights which inputs most affect the target variable, strength of influence further clarifies the magnitude of these effects along each arc, providing a complementary layer of causal interpretation.
To further quantify the directional dependencies within the BNs, we used the Strength of Influence feature available in GeNIe.  This analysis captures the extent to which a parent node affects the probability distribution of its child node, and measures the average change in the child’s posterior distribution when the parent transitions between its states where higher values indicate stronger influence. This complements the structural analysis by prioritizing relationships not only by presence but also by magnitude of effect. In the visualisations, the thickness of each arc reflects the corresponding strength of influence, and greater thickness indicates a stronger relationship between the connected nodes.

The strength of influence analysis highlights distinct roles for each cryptocurrency within its respective BN. As shown in Figure~\ref{fig:BTC_sen}, Bitcoin exerts a stronger influence on tweet activity than it receives from financial assets. This is evident from the thicker outgoing arc to Tweet No. compared to the thinner arc from MSCI to Bitcoin. Moreover, Binance Coin (Figure~\ref{fig:BNB_sen}) receives input from MSCI and has connections to both Tweet No. and the S\&P 500. The arc toward the S\&P 500 is thicker than the one to Tweet No., which may suggest a stronger connection to financial variables than to social sentiment. Moreover, in Figure~\ref{fig:ETH_sen}, the arc from Ethereum to Tweet No. is clearly visible and thicker than any other in its network, indicating that Ethereum may play a stronger role in shaping sentiment compared to other cryptocurrencies.

Both Ripple and Litecoin exhibit low influence strengths with their connected nodes. Although they are structurally linked to financial variables such as MSCI and the S\&P 500, the arcs are narrow and less prominent (Figures~\ref{fig:LTC_sen} and \ref{fig:XRP_sen}), suggesting that these inputs have limited impact under the current model.



\section{Conclusions and Future Work} \label{Conclusions}

This study investigates the drivers of cryptocurrency price movements using Bayesian networks, modelling relationships among six major cryptocurrencies, five traditional financial assets, and daily tweet volumes. BNs are selected for their ability to uncover causal dependencies and support probabilistic reasoning. Given the continuous nature of price data, the study systematically evaluates multiple discretisation strategies and proposes a structured BN construction pipeline. Using this pipeline, nine BNs are constructed for each cryptocurrency based on different discretisation methods and bin settings. Their predictive performance is evaluated using four metrics to identify the most accurate models and uncover key price-driving factors.

From a total of 54 constructed BNs, the top-performing BN for each cryptocurrency is selected based on a composite score. Results indicate that equal interval discretisation with two bins yields the most accurate and robust performance across metrics. Analysis of these metrics reveals that different cryptocurrencies respond differently to the same model configurations, underscoring the importance of coin-specific modelling. This suggests that a single investment strategy is unlikely to be effective across all cryptocurrencies due to their distinct behaviours and varying sensitivities to external drivers. Notably, equal interval discretisation with two bins emerged as the most robust configuration overall, though its effectiveness varied across coins.



In addition, post-hoc analyses, including sensitivity analysis and strength of influence evaluation, highlight which variables exert the greatest impact on price dynamics. These methods enhance model transparency and help investors and researchers prioritise the most influential economic and social factors in the cryptocurrency market. The findings reveal that, although there is a connection between social media activity and cryptocurrency prices, the direction of influence often flows from price movement to social media engagement, suggesting that price changes tend to drive public discourse rather than result from it. Furthermore, the relationship between traditional financial assets and cryptocurrency prices is coin-dependent, reinforcing the need for tailored, coin-specific modelling approaches.

Building on these insights, several future directions are suggested. First, major geopolitical and macroeconomic events, such as China’s crypto bans~\citep{shahzad2018empirical}, the Russia–Ukraine conflict, and COVID-19~\citep{albulescu2021covid}
, warrant deeper analysis for their effects on cryptocurrencies and global markets, particularly in detecting regime shifts. Second, incorporating expert knowledge into BN learning could enhance causal inference and model interpretability, especially when assessing the impact of influential figures~\citep{ante2021elon}. Third, as an important direction, to improve the robustness of causal inference in dynamic markets, a promising direction is to extend the current approach using dynamic Bayesian networks to account for temporal dependencies among influencing price factors~\cite{amirzadeh2023dynamic}.
Lastly, examining internal blockchain data, such as transactions, hash rates,  and integrating sentiment or search trends, including Google Trends, may improve prediction and support more adaptive investment strategies.


\section*{Declarations}
\subsection*{Conflict of interest}
The authors have no competing interests to declare that are relevant to the content of this article.
\subsection*{Funding}
No funding was received to assist with the preparation of this manuscript.

\subsection*{Contributions}
All authors contributed equally to this study.

\subsection*{Ethical approval}
This article does not contain any studies with human participants and Animals performed by authors.

\subsection*{Data availability }
Data were derived from the following resource available in the public domain:
https://finance.yahoo.com/


\newpage
\begin{appendix}

\section{Descriptive Analysis of Data Frames}\label{Descriptive}

In this section, we provide a statistical analysis of the cryptocurrency datasets used in this study, focusing on price and tweet volume across six major coins over the period from January 2018 to April 2023.

\subsection{Summary Statistics of Prices and Tweet Volumes}\label{Prices_Tweet_stats}

Tables~\ref{tab:coin_describtion} summarise the descriptive statistics of the daily closing prices after merging with traditional asset data and tweet volume. The dataset was constructed by aligning daily records from all available sources. The columns are self-explanatory, and the {\it No. Obs.} column indicates the number of valid observations in the merged dataset. While missing values in the price and macroeconomic variables were negligible, the Twitter data exhibited more frequent gaps. Consequently, the variation in observation counts across assets is primarily due to missing entries in the tweet data.

\begin{table*}[h]
 \centering
 \caption{Summary statistics for the daily close prices of cryptocurrencies.}
 \label{tab:coin_describtion}
 \begin{adjustbox}{max width=\textwidth}
 \begin{tabular}{lllllll}
 \hline
 \textbf{Altcoin} & \textbf{Mean} & \textbf{Std. Dev.} & \textbf{Min.} & \textbf{Median} & \textbf{Max.} & \textbf{No. Obs.} \\
 \hline
 Bitcoin       & 20664.35 & 16771.39 & 3242.48  & 11570.15 & 67566.83 & 1282 \\
 Binance Coin  & 161.88   & 183.92   & 4.53     & 29.23    & 675.68   & 1264 \\
 Ethereum      & 1169.65  & 1173.12  & 84.31    & 586.01   & 4151.87  & 1281 \\
 Litecoin      & 96.91    & 56.65    & 23.46    & 75.54    & 250.49   & 1281 \\
 Ripple        & 0.49     & 0.28     & 0.14     & 0.39     & 1.15     & 1277 \\
 Tether        & 1.0010   & 0.0022   & 0.9968   & 1.0004   & 1.0053   & 1307 \\
 \hline
 \end{tabular}
 \end{adjustbox}
\end{table*}

Tables~\ref{tab:coin_describtion} and~\ref{tab:tweet_no_describtion} show the descriptive statistics of daily tweet counts for each cryptocurrency. 
The table shows notable differences in tweet activity across altcoins. Bitcoin has the highest average and variability in tweet volume, indicating its central role in market discussions. Ethereum and Ripple also receive considerable attention, reflected in their relatively high tweet counts. In contrast, Binance Coin and Litecoin have lower and more stable tweet volumes, suggesting more limited or consistent public engagement. These differences highlight the varying levels of market attention each cryptocurrency attracts on social media.
\begin{table*}[h]
 \centering
 \caption{Summary statistics for daily tweet volumes related to altcoins.}
 \label{tab:tweet_no_describtion}
 \begin{adjustbox}{max width=\textwidth}
 \begin{tabular}{llllll}
 \hline
 \textbf{Altcoin} & \textbf{Mean} & \textbf{Std. Dev.} & \textbf{Min.} & \textbf{Median} & \textbf{Max.} \\
 \hline
 Bitcoin       & 75916.23  & 56546.05  & 445     & 50454.00  & 252186 \\
 Binance Coin  & 95.75     & 89.95     & 1       & 63.00     & 280 \\
 Ethereum      & 18146.17  & 14133.48  & 2418    & 14007.00  & 60569 \\
 Litecoin      & 1388.96   & 772.13    & 360     & 1154.00   & 3175 \\
 Ripple        & 12675.62  & 9730.45   & 2362    & 8479.00   & 35596 \\
 \hline
 \end{tabular}
 \end{adjustbox}
\end{table*}

\begin{figure*}[h]
  \centering
\includegraphics[width=\linewidth,height=\textheight,keepaspectratio]{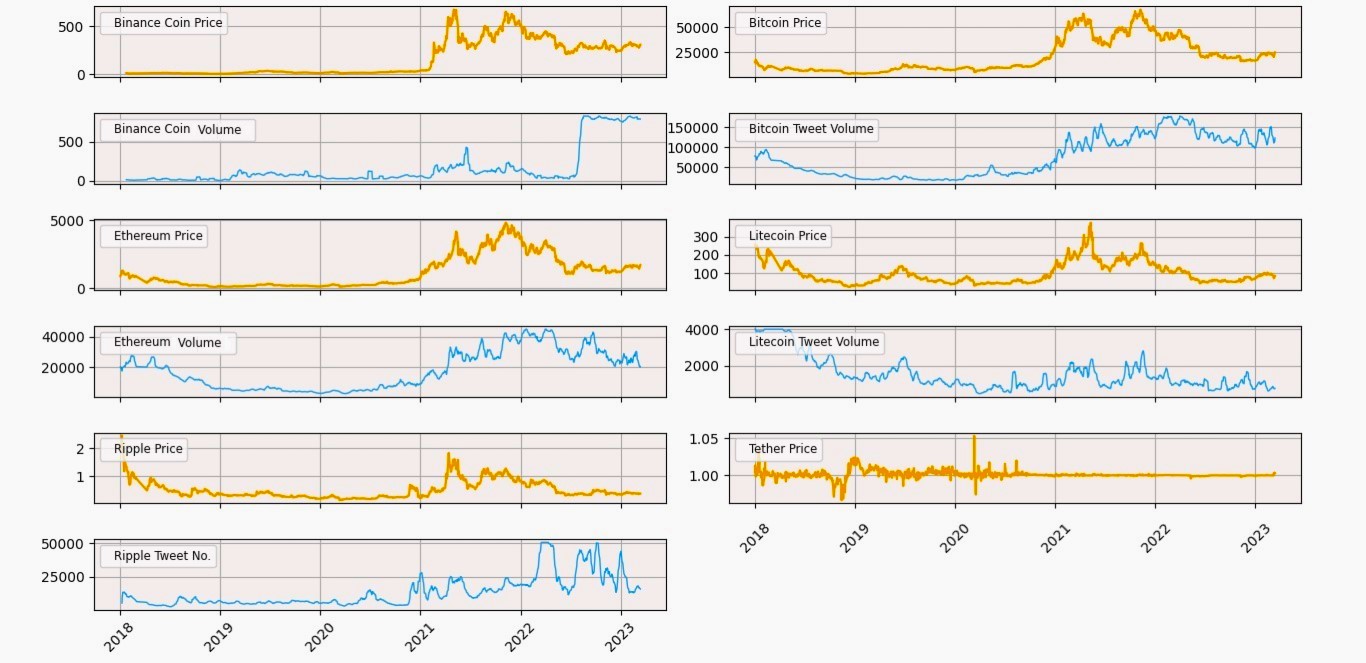}
\caption{Daily close prices (orange) and tweet volumes (blue) for each altcoin from 2018 to 2023. To reduce the impact of extreme outliers and enhance readability, values above the 90th percentile are removed.}
  \label{fig:tweet_price_all}
\end{figure*}

To better illustrate the relationship between price trends and social media activity, Figure~\ref{fig:tweet_price_all} shows the daily prices and tweet volumes for each cryptocurrency. As seen in the figure, price fluctuations are most noticeable during 2021, coinciding with a surge in tweet activity. This period reflects a market momentum and investor attention, possibly linked to broader market regime shifts such as responses to COVID-19. Notably, Bitcoin and Ethereum exhibit consistently high tweet volumes throughout the period, suggesting ongoing investor engagement. In contrast, altcoins such as Binance Coin and Ripple display more spikes in tweet activity, typically aligned with sharp price movements. This indicates that discourse around some assets is more reactive and event-driven, whereas others maintain broader and more continuous public attention.

\subsection{Distributional Properties and Outlier Detection}\label{Outlier}

The descriptive statistics in Table~\ref{tab:coin_stats_descriptive} presents summary statistics on the distributional characteristics and outlier counts for each cryptocurrency. As can be seen, Bitcoin and Binance Coin exhibit moderate right skewness with light tails and no detected outliers. Ethereum shows a right-skewed distribution with moderate tail thickness and a moderate number of outliers, reflecting occasional deviations from central trends. Litecoin and Ripple display stronger positive skewness and heavier tails, consistent with more volatile behaviour and higher frequencies of extreme price changes. Ripple, in particular, shows the greatest asymmetry and volatility, as indicated by its high number of outliers. Tether, despite having relatively low skewness, exhibits extremely high kurtosis due to its narrow price range, where even small fluctuations are statistically identified as outliers.

\begin{table*}[!ht]
\centering
\caption{Descriptive statistics for each cryptocurrency: number of outliers, skewness, and kurtosis.}
\label{tab:coin_stats_descriptive}
\begin{adjustbox}{max width=\textwidth}
\begin{tabular}{lccc}
\hline
\textbf{Coin} & \textbf{Outliers} & \textbf{Skewness} & \textbf{Kurtosis} \\
\hline
Bitcoin       & 0   & 1.00 & -0.27 \\
Binance Coin  & 0   & 0.82 & -0.65 \\
Ethereum      & 29  & 1.10 &  0.16 \\
Litecoin      & 32  & 1.34 &  1.66 \\
Ripple        & 62  & 1.69 &  3.32 \\
Tether        & 223 & 0.73 & 16.97 \\
\hline
\end{tabular}
\end{adjustbox}
\end{table*}

\subsection{Stationarity Evaluation for Time Series Modelling}\label{adf}

To assess the stationarity of the data for time series modelling, we applied both the Augmented Dickey-Fuller (ADF) and Kwiatkowski–Phillips–Schmidt–Shin (KPSS) tests. A time series is considered stationary if it yields a p-value below 0.05 in the ADF test and above 0.05 in the KPSS test. As shown in Table~\ref{tab:stationarity_tests}, most cryptocurrency price series failed to satisfy both tests.

\begin{table*}[h]
\centering
\caption{Results of ADF and KPSS tests for stationarity.}
\label{tab:stationarity_tests}
\begin{adjustbox}{max width=\textwidth}
\begin{tabular}{lcccc}
\toprule
\textbf{Altcoin} & \textbf{ADF Test Statistic} & \textbf{ADF p-value} & \textbf{KPSS Test Statistic} & \textbf{KPSS p-value} \\
\midrule
Bitcoin       & -1.7294 & 0.4160 & 3.0818 & 0.0100 \\
Binance Coin  & -1.8062 & 0.3774 & 3.9603 & 0.0100 \\
Ethereum      & -1.5446 & 0.5113 & 3.2603 & 0.0100 \\
Litecoin      & -2.9096 & 0.0442 & 0.6457 & 0.0185 \\
Ripple        & -2.9066 & 0.0446 & 0.8434 & 0.0100 \\
Tether        & -5.0250 & 0.0000 & 1.2232 & 0.0100 \\
\bottomrule
\end{tabular}
\end{adjustbox}
\end{table*}

Among the cryptocurrencies, only Litecoin and Ripple partially satisfied the stationarity criteria by rejecting the ADF null hypothesis but failing the KPSS test, suggesting weak stationarity. Tether, being a stablecoin, was the only asset that met both criteria and exhibited stationarity without transformation. Consequently, all other series were transformed using first-order differencing prior to modeling to ensure stationarity.

\end{appendix}

\end{document}